\documentclass[%
reprint,
superscriptaddress,
twocolumn,
amsmath,
amssymb,
aps,
prl
]{revtex4-1}

\usepackage{graphicx}  
\usepackage{dcolumn}  
\usepackage{bm}            
\usepackage{bbold}
\usepackage{amsmath}
\usepackage{xfrac}
\usepackage{xcolor}
\usepackage[colorlinks=true, citecolor=red]{hyperref}   
\usepackage{hhline}
\usepackage{diagbox}
\usepackage{siunitx}
\usepackage{enumitem}
\usepackage[normalem]{ulem}

\newcommand{\colorrev}{\color{black}}

\newcommand{\ac}{\alpha_c\!}
\newcommand{\av}{\alpha_v\!}

\begin{document}
	
	\title{Long-Lived Topological Flatband Excitons in Semiconductor Moir\'e Heterostructures:\\
		  A  Bosonic Kane-Mele Model Platform
    }

	\author{Ming Xie}
     		\email{mingxie@umd.edu}
	  	    \affiliation{Condensed Matter Theory Center, Department of Physics, 
			                University of Maryland, College Park, Maryland 20742, USA}
	\author{Mohammad Hafezi}
		\affiliation{Joint Quantum Institute, Department of Physics, 
							University of Maryland, College Park, Maryland 20742, USA}
	\author{Sankar Das Sarma}
    	\affiliation{Condensed Matter Theory Center, Department of Physics, 
    		                University of Maryland, College Park, Maryland 20742, USA}
    	                \affiliation{Joint Quantum Institute, Department of Physics, 
    	                	University of Maryland, College Park, Maryland 20742, USA}
    	                
	\date{\today}
	
	\begin{abstract}
		
		Moir\'e superlattices based on two-dimensional transition metal dichalcogenides (TMDs) 
		have emerged as a highly versatile and fruitful platform for exploring correlated topological 
		electronic phases. One of the most remarkable examples is the recently discovered
		fractional quantum anomalous Hall effect (FQAHE) under zero magnetic field. 		
		Here we propose a minimal structure that hosts long-lived excitons---a ubiquitous 
		bosonic excitation in TMD semiconductors---with narrow topological bosonic bands.
		The nontrivial exciton topology originates from hybridization of moir\'e interlayer excitons
		and is tunable by controlling twist angle and electric field. 
		At small twist angle, the lowest exciton bands are isolated from higher energy bands
		and provide a solid-state realization of the bosonic Kane-Mele model with topological flatbands,
		which could potentially support the bosonic version of FQAHE.

	\end{abstract}
	
	\maketitle

\emph{Introduction}---\noindent
Owing to their narrow bandwidths and intertwined electronic wave functions,
moir\'e superlattices formed from overlaying two-dimensional transition metal dichalcogenide (TMD) semiconductors 
have become a unique crossroad where strong electron correlations and nontrivial topology meet,
under unprecedented controllability.
In TMD superlattices where band topology is trivial, a plethora of strongly correlated 
phenomena, such as  Mott insulators, generalized Wigner crystals, and 
metal-insulator transitions, have been identified
\cite{MakHubbard2020, Dean2020, WignerCrystal, Shan2020, LeRoy, StripePhase,  
	MakMott2021, QuantumCriticality, Pasupathy2021, Cui2021, ShanMIT, QuantumCriticality, HeavyFermion}.
Recently, in TMD superlattices featuring topological bands, experimental studies 
observed correlated phases with nontrivial topological characteristics
\cite{MakQAHE2021, Cai2023, Zeng2023, Park2023, Fan2023, Feldman2023, Tao2024, MakQSHMoTe2, MakQSHWSe2},
including both integer and fractional quantum anomalous Hall states.
The latter, also referred to as fractional Chern insulators, are observed for the first time
in experiments after being proposed for over a decade
\cite{Tang2011, Sun2011, Neupert2011, Regnault2011, Sheng2011}.
These observations establish TMD moir\'e superlattices 
as promising candidates for solid-state fermionic quantum simulations.

The drastic influence of the long-period moir\'e pattern extends to 
excitons---tightly bound electron-hole pair excitations---in TMD moir\'e superlattices,
rendering their localization on moir\'e lattices
\cite{Xu2019, Li2019, Feng2019, Tartakovskii2019, Lui2021, Fengtrilayer2022, Deng2021,
	TopoIntraX, Yao2017, WuInterXPRB, MoireDiracCone, YaoPRX2021}.
Remarkably, recent experiments discovered a correlated incompressible state of excitons 
\cite{Jin2023, Hafezi2023, Shi2021, XuLadder, Shi2024, Jin2024}, 
or a bosonic Mott insulator, in TMD moir\'e heterobilayers,
breaking new ground for exploring many-body states of bosons.
The moir\'e modulation in these superlattices leads to localized excitons
 seating on a triangular lattice with strong on-site interactions, 
effectively simulating the Bose-Hubbard model. 
The emergence of localized bosonic lattices can be understood as 
the formation of trivial excitonic moir\'e bands with narrow bandwidths.
One intriguing question arises: 
Is it possible to form topological moiré bands for excitons, 
thereby paving the way to achieve bosonic correlated topological phases,
such as bosonic fractional Chern insulators?

It was understood that excitons in the $+K$ and $-K$ valleys of TMD monolayers have nonzero Berry curvatures 
due to valley-momentum coupling induced by exchange interactions
\cite{FWuXband, Louie2015, MonoDiracCone, TingCao2012}.
Based on this, Wu and coworkers \cite{TopoIntraX} showed that, when such intralayer exciton is subjected 
to a periodic moir\'e potential, the resulting low-energy moir\'e exciton bands can acquire 
a definite Chern number, provided an effective valley Zeeman field is included.
However, one critical issue that remains unaddressed in this scenario is the short lifetime of excitons,  
which is detrimental for experimental realization of many-body states of excitons.
Intralayer excitons are known to have large optical dipole moment
responsible for their short recombination lifetime;
they decay rapidly before a quasiequilibrium population of excitons can be established.
The fact that the strength of the valley-momentum coupling is proportional to the optical dipole
moment places topology and long lifetime at odds with each other.

In this Letter, we propose a minimal TMD moir\'e heterostructure capable of 
supporting excitons that are both long-lived and topological.
It is based on interlayer moir\'e excitons whose optical dipole moments nearly vanish
due to layer separation of their constituent electron and hole.
The topology of the exciton moir\'e bands originates from spatially varying hybridization 
of interlayer excitons situated in different layers of the moir\'e heterostructure.
This mechanism does not involve the exchange-induced exciton Berry curvature,
thereby avoiding the conflict between topology and long lifetime.
We develop an effective bosonic continuum model for the interlayer moir\'e excitons.
The valley-projected exciton band structure features a rich set of bosonic topological bands
with opposite Chern numbers for opposite valley pseudospins.
We find that, as the twist angle is varied, 
the bandwidth of the lowest exciton band exhibits a minimum 
at a ``magic" angle.
At small twist angles, the pair of  isolated lowest-energy exciton bands
provides a first solid-state realization of the bosonic Kane-Mele model with nearly flat topological bands.

\begin{figure}[t!]
	\includegraphics[width=0.48\textwidth]{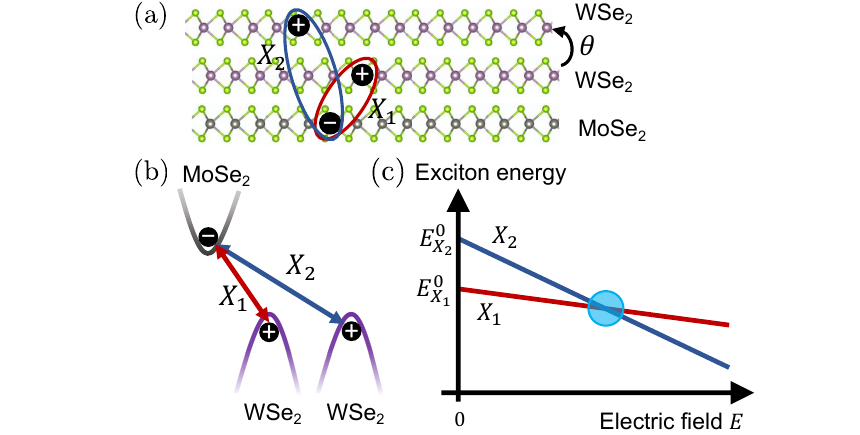}
	\caption{\label{trilayer}(a) Schematic illustration of the proposed moir\'e heterostructure. The WSe$_2$ layers are twisted relative to each other by an angle $\theta$ while the MoSe$_2$ layer is aligned with the middle WSe$_2$ layer. (b) The band diagram hosting the two low-energy interlayer excitons, $X_1$ and $X_2$, illustrated in (a). (c) Energy of the $X_1$ and $X_2$ excitons as a function of the vertical electric field.
	}
\end{figure}

\emph{Moir\'e heterostructure}---\noindent
The proposed structure consists of a twisted $MX_2$ ($tMX_2$) homobilayer 
stacked on top of another $M'X_2$ monolayer, with rotation alignment between 
the interfacing $MX_2$ and $M'X_2$ layers, as illustrated in Fig.~\ref{trilayer}(a),
where $M, M'$ represent different transition metal atoms, W or Mo, 
and $X$ represents a S, Se, or Te atom.
We have chosen TMD layers with a common chalcogen element ($X$), 
{\colorrev for their negligible mismatch in lattice constant  \cite{interfacenote}},
to avoid unimportant complications due to the formation of a second moir\'e pattern at the heterointerface. 
(Our theory also applies, in this regard, if a thin $h$BN spacer is added.) 
As will become clear below, it is the $tMX_2$ that provides the moir\'e-modulated hybridization critical to our theory.

In this work, we focus on the tWSe$_2$/MoSe$_2$ heterostructure as an example \cite{footnote}, 
which features a type-II band alignment
with a momentum-direct interlayer band gap between
the conduction band of the MoSe$_2$ and
the valence bands of the tWSe$_2$ at the $\pm K$ valleys
[Fig.~\ref{trilayer}(b)].
The lowest-energy excitons are the intravalley interlayer excitons
formed by electrons in the MoSe$_2$ layer and holes in the middle and top WSe$_2$ layers,
labeled as $X_1$ and $X_2$, respectively [shown in Fig.~\ref{trilayer}(a)].
Because of their different electron-hole vertical separations,
$X_1$ and $X_2$ excitons have different binding energies.
Their energies, $E_{X_l}=E^0_{X_l}-elV_b/2$ (with $e>0$), 
can be tuned by a perpendicular electric field $E$ as shown in Fig.~\ref{trilayer}(c),
where $l=1,2$, $V_b=2\epsilon Ed$ is the bias potential across the trilayer, $d$ is interlayer spacing,
and $E^0_{X_l}$ is the exciton energy at $E=0$.
As we demonstrate below, interesting physics occurs when $X_1$ and $X_2$ are close in energy, and hybridize with each other [highlighted in cyan in Fig.~\ref{trilayer}(c)].

The electron-hole two-body problem in the moir\'e superlattice can be simplified,
taking advantage of the facts that the exciton binding energy ($\sim$100 meV) 
is much larger than the interlayer hybridization strength ($\sim10\text{--}20$ meV),
and that the binding energy variation in the superlattice potential
is generally smaller than the band gap variation \cite{WuInterXPRB,YaoPRX2021}.
This allows us to treat the exciton problem and the effect of the superlattice
separately in a two-step process.
We start by solving the Bethe-Salpeter equations (BSEs) \cite{TopoIntraX,Louie2015, WuInterXPRB} 
in the absence of the moir\'e potential and interlayer hybridization:
\begin{align}
		\begin{split}
	&(E_{\tau,c,{\textbf{k}+\ac\textbf{Q}}}-E_{\tau,vl,{\textbf{k}-\av\textbf{Q}}})
	A^{\tau}_{S,l,{\textbf{Q}}}(\textbf{k}) \vspace{5em}
	\\&\negmedspace+\negmedspace\sum_{\textbf{k}'} \langle\tau l\textbf{k}\textbf{Q} |V|\tau l\textbf{k}'\textbf{Q} \rangle A^{\tau}_{S,l,{\textbf{Q}}}(\textbf{k}') 
	\negmedspace=\negmedspace\mathcal{E}^{\tau}_{S,l}(\textbf{Q}) A^{\tau}_{S,l,{\textbf{Q}}}(\textbf{k}),
		\end{split}
\end{align}
where $\textbf{Q}$ and $\textbf{k}$ are the center-of-mass (c.m.) 
and relative momentum of the electron-hole state,
$|\tau l\textbf{k}\textbf{Q} \rangle \equiv \hat{c}^{\dagger}_{\tau c\textbf{k}+\ac\textbf{Q}} 
\hat{c}_{\tau vl\textbf{k}-\av\textbf{Q}}|0\rangle$.~$\hat{c}^{\dagger}$($\hat{c}$) 
is the electron creation (annihilation) operator, and $|0\rangle$ is the charge neutrality state.
$\alpha_{c(v)}=m^*_{c(v)}/M$ with $M=m^*_{c}+m^*_{v}$ the exciton effective mass.
$\tau=\pm K$ is the valley index, and $l=1,2$ denotes 
the middle ($l=1$) and the top ($l=2$) WSe$_2$ layers.
$S$ is the exciton band index.
We adopted the usual convention that $\textbf{k}$ is measured from the $\pm K$ point of each layer.
$\langle \tau l\textbf{k}\textbf{Q} |V|\tau l\textbf{k}'\textbf{Q} \rangle$ 
represents the direct interlayer Coulomb interaction matrix
(see Supplemental Material (SM) for details \cite{SM}).

The lowest-energy solution ($S=0$) of each BSE is the interlayer exciton state $X_l$ with 
energy $\mathcal{E}^{\tau}_l(\textbf{Q})$ and wave function:
\begin{align}
	|X_l(\textbf{Q})\rangle_{\tau} =\sum_{\textbf{k}} A^{\tau}_{l{\textbf{Q}}}(\textbf{k})\hat{c}^{\dagger}_{\tau
	 c\textbf{k}+\ac\textbf{Q}} \hat{c}_{\tau vl\textbf{k}-\av\textbf{Q}}|0\rangle.
 \label{Xlbase}
\end{align}
The internal wave function $A^{\tau}_{l{\textbf{Q}}}(\textbf{k})$
is nearly independent of $\textbf{Q}$
and isotropic in $\textbf{k}$, reminiscent of the $1s$ exciton wave function.
(We have dropped the $S$ index for convenience.)
$\mathcal{E}^{\tau}_l(\textbf{Q})\approx E_{X_l} + \hbar^2\textbf{Q}^2/M$ at small $\textbf{Q}$.
The energy difference $\Delta E^0\equiv E^0_{X_2}-E^0_{X_1}$ is about 50 meV.
We note that a more accurate treatment of the exciton problem 
could be achieved using the \textit{ab initio} GW-BSE approach \cite{Louie2015, Louie2000},
which, however, falls outside the focus and scope of the this work.

\begin{figure}[t!]
	\includegraphics[width=0.43\textwidth]{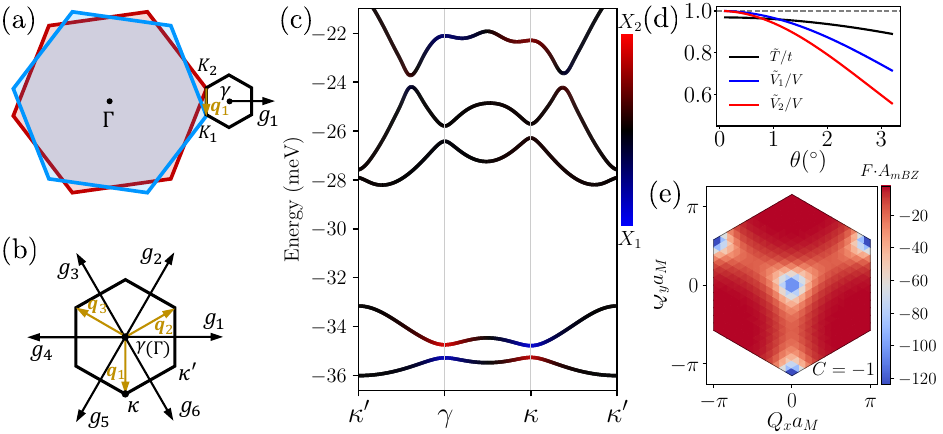}
	\caption{\label{excitonband}
		(a) Rotated Brillouin zones of the middle (cyan) and top (red) WSe$_2$ layers
		 and the moir\'e Brillouin zone (black) of the tWSe$_2$.
		(b) Exciton moir\'e Brillouin zone with its center ($\gamma$ point)
		set at the origin $\Gamma$ of momentum space. 
		(c) Band structure moir\'e excitons at $\theta=1.5^{\circ}$ and $eV_b-2\Delta E_0=4.2$ meV.
		Blue (red) line color indicates the fraction of the $X_1$($X_2$) component.
		(d) $\tilde{V}_l/V$ and $\tilde{T}/t$ as a function of the twist angle.
		(e) Berry curvature of the lowest exciton band with Chern number $C_0=-1$.
	}
\end{figure}

\emph{Exciton moir\'e  Hamiltonian}---\noindent
We now derive a continuum model Hamiltonian for the interlayer excitons, taking
into account the effect of the moir\'e superlattice modulation and interlayer hybridization.
Since the valence bands are situated in the tWSe$_2$ superlattice, 
the hole component of the excitons experience a moir\'e potential modulation
\cite{TopoIntraX}.
The moir\'e potential takes the form \cite{WumoireHubbard, WumoireTI}
\begin{align}
	U_l(\bm{r})= 2V\sum_{i=1,3,5} \cos\negmedspace\big(\bm{g}_{i}\cdot\bm{r}+s_{i,l}\phi\big),
\end{align}
where $V$ and $\phi$ are the amplitude and phase parameters, respectively, and $s_{i,l}=(-1)^{i+l-1}$.
$\bm{g}_1={4\pi}/{\sqrt{3}a_M}(1, 0)$ and $\bm{g}_i=(\hat{\mathcal{R}}_{\pi/3})^{i-1}\bm{g}_1$ 
are moir\'e reciprocal lattice vectors 
where $\hat{\mathcal{R}}_{\pi/3}$ is counterclockwise rotation around the $z$ axis by $\pi/3$ [Fig.~\ref{excitonband}(a)].
By Fourier transforming $U_l(\bm{r})$  and projecting it to the exciton basis in Eq.~(\ref{Xlbase}),
we obtain the exciton moir\'e potential:
\begin{align}
	\label{Hintra}
	\mathcal{U}^+_{l,\textbf{Q},\textbf{Q}'} = -\tilde{V}_l\sum_{i=1}^6  \exp\negmedspace\big(is_{i,l}\phi\big) 
	\delta_{\textbf{Q}-\bm{g}_i, \textbf{Q}'},
\end{align}
where the superscript ``$+$" indicates valley,
and the minus sign reflects that excitons experience an opposite sign of moir\'e potential compared to that of the valence electrons.
The amplitude $\tilde{V}_l= V\sum_{\bm{k}} A^{+}_{l,0}(\bm{k}) A^{+}_{l,0}(\bm{k}+\ac\,\bm{g}_1)$
depends on the overlap of exciton internal wave functions with a relative shift $\ac\,\bm{g}_1$ 
(see SM \cite{SM} for details) and decreases with increasing twist angle, 
 as shown in Fig.~\ref{excitonband}(d).
The delta function ensures the conservation of exciton c.m. momentum $\textbf{Q}$ 
up to a moire reciprocal vector.

The $X_1$ and $X_2$ excitons hybridize with each other via
 the interlayer hybridization between the valence bands of the tWSe$_2$,
which varies periodically in real space leading to nontrivial layer pseudospin winding 
\cite{WumoireTI, Pan2020, FuMagicAngle, Yu2020, Zhai2020}. 
In the lowest-order harmonic approximation, it takes the form for valley $+K$
\begin{align}
		T^+(\bm{r}) = t\sum_{n}e^{i\bm{q}_n\cdot\bm{r}} ,
\end{align}
where $t$ is the tunneling strength. 
$\textbf{q}_1=\textbf{K}_1-\textbf{K}_2$ is the momentum shift 
between the $K$ points in the two WSe$_2$ layers
and $\textbf{q}_n$ is related to $\textbf{q}_1$ by threefold rotation
$\textbf{q}_n=(\hat{\mathcal{R}}_{2\pi/3})^{n-1}\textbf{q}_1$ [Fig.~\ref{excitonband}(a)].
Similar to Eq.~\ref{Hintra}, we obtain the hybridization Hamiltonian:
\begin{align}
	\label{Hinter}
	\mathcal{T}^+_{\textbf{Q},\textbf{Q}'} = t\sum_{n=1}^3\sum_{\textbf{k}}\negmedspace A^{+*}_{1,{\textbf{Q}}}(\textbf{k})A^{+}_{2,{\textbf{Q}'}}\big(\textbf{k}\negmedspace-\negmedspace
	\alpha_c\bm{q}_n\big) \delta_{\textbf{Q}+\textbf{q}_n,\textbf{Q}'},
\end{align}
where the amplitude for the three hopping processes 
$\tilde{T}_n\equiv t\sum_{\textbf{k}} A^{+*}_{1,{\textbf{Q}}}(\textbf{k})A^+_{2,{\textbf{Q}+\textbf{q}_n}}
(\textbf{k}\negmedspace-\!\ac\,\textbf{q}_n)$,
are equal $\tilde{T}_1=\tilde{T}_2=\tilde{T}_3\equiv \tilde{T}$ because of the $C_3$ symmetry (around the $z$ axis).
Similar to $\tilde{V}_l/V$, $\tilde{T}/t$ decreases monotonically with increasing $\theta$ 
[shown in Fig.~\ref{excitonband}(d)].
It should be emphasized that the c.m. momentum of $X_2$ exciton is measured from $\bm{\kappa}=\textbf{K}_1-\textbf{K}_2$ instead of from $\bm{\gamma}$ for the $X_1$ exciton.

Combining Eqs.~(\ref{Hintra}) and (\ref{Hinter}),
we arrive at the full exciton moir\'e Hamiltonian (for valley $\tau=+K$)
in the basis $\{|X_1\rangle_+, |X_2\rangle_+\}$:
\begin{align}
	\label{HamX}
	\mathcal{H}^+_{\textbf{Q},\textbf{Q}'} \negmedspace=\negmedspace 
	\begin{pmatrix}
		\mathcal{E}_1(\textbf{Q}) \delta_{\textbf{Q},\textbf{Q}'}\negmedspace+ \mathcal{U}^+_{1,\textbf{Q},\textbf{Q}'} 
		& \mathcal{T}^+_{\textbf{Q},\textbf{Q}'} \\
		[\mathcal{T}^{+}]^\dagger_{\textbf{Q},\textbf{Q}'} & 
		\mathcal{E}_2(\textbf{Q}) \delta_{\textbf{Q},\textbf{Q}'}\negmedspace+ \mathcal{U}^+_{2,\textbf{Q},\textbf{Q}'}
	\end{pmatrix}\negmedspace,
\end{align}
which respects the $C_3$ symmetry and is related to $\mathcal{H}^-$
by time-reversal symmetry $\mathcal{T}$.
We choose the exciton moir\'e Brillouin zone (mBZ) 
whose center ($\gamma$ point) sits at the origin of  c.m. momentum $\Gamma$, 
as illustrated in Fig.~\ref{excitonband}(b),
and numerically diagonalize $\mathcal{H}^\pm$ to obtain 
the exciton moir\'e band structure \cite{SM}. 

\emph{Exciton topological flatbands}---\noindent
The valley-projected ($\tau\negmedspace=\negmedspace+K$) exciton moir\'e band structure is 
shown in Fig.~\ref{excitonband}(c) at a representative twist angle $\theta=1.5^{\circ}$
and $eV_b-2\Delta E_0=4.2$ meV.
The band structure features two narrow low-energy bands
 isolated from higher-energy bands. 
A gap opens up between the two as a result of the hybridization between the $X_1$ and $X_2$ excitons.
The Berry curvature of the lowest band, illustrated in Fig.~\ref{excitonband}(e),
exhibit strong amplitudes at high symmetry points $\gamma$ and $\kappa$,
reminiscent of a band inversion that occurs at these points (see SM \cite{SM}).
The Chern numbers of the two lowest bands are $C_0=-1$ and $C_1=1$,
indicating that a pair of time-reversal symmetry protected bosonic helical edge states exist inside their gap.

\begin{figure}[t!]
	\includegraphics[width=0.47\textwidth]{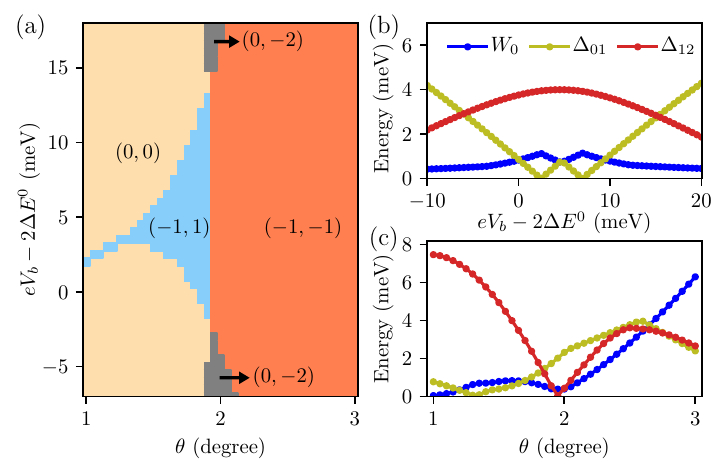}
	\centering
	\caption{\label{phasediagram}
		(a) Topological phase diagram of the moir\'e exciton bands.
		The numbers in the parentheses are $(C_0, C_1)$, the Chern number of the two lowest bands.
		(b) Bandwidth and energy gap as a function of $eV_b$ at $\theta=1.6^{\circ}$. 
		$W_0$ is the bandwidth of the lowest ($n=0$) moir\'e band.
		$\Delta_{01}$ ($\Delta_{12}$) is the global gap between the $n=0$ and $1$ ($n=1$ and $2$) bands.
		(c) Bandwidth and energy gaps as a function of $\theta$ at fixed $eV_b-2\Delta E^0=5$ meV.
	}
\end{figure}

Upon varying twist angle and displacement field, 
the exciton bands display a rich topological phase diagram
as illustrated in Fig.~\ref{phasediagram}(a).
As $eV_b$ is tuned closer to $2\Delta E^0$, 
the band inversion between the first and second moir\'e bands occurs, 
resulting in a topological transition from the trivial phase with 
$(C_0, C_1)\negmedspace=\negmedspace(0,0)$
to the topological phase with $(C_0, C_1)\negmedspace=\negmedspace(-1,+1)$.
The gap between the two lowest bands $\Delta_{01}$ closes at the transition point as shown in Fig.~\ref{phasediagram}(b).
For small twist angle $\theta<1.4^{\circ}$,
the displacement field range for the inverted region is narrow because of the small bandwidths.
As $\theta$ increases,
the $n=1$ band intersects the higher ($n=2$) band at around $\theta=2^{\circ}$ 
[where $\Delta_{12}$ vanishes as shown in Fig.~\ref{phasediagram}(c)],
leading to a change in the Chern number, from $C_1=+1$ to $C_1=-1$,
while $C_0$ remains unchanged.
This intersection also leads to the transition from the 
$(0,0)$ to $(0,-2)$ phase at large $|eV_b-2\Delta E^0|$.
Interestingly, the bandwidth $W_0$ of the lowest band exhibits a minimum 
close to zero at around $\theta_m=2^{\circ}$,
marking the first emergence of  a``magic angle" for a bosonic topological band.
{\colorrev Exciton-exciton interactions, estimated to be around 30--80 meV \cite{Jin2023, Hafezi2023, Shi2021, XuLadder, Shi2024, Jin2024, UexPRB}, dominate over the bandwidth at small twist angles.}

\begin{figure}[t!]
	\includegraphics[width=0.42\textwidth]{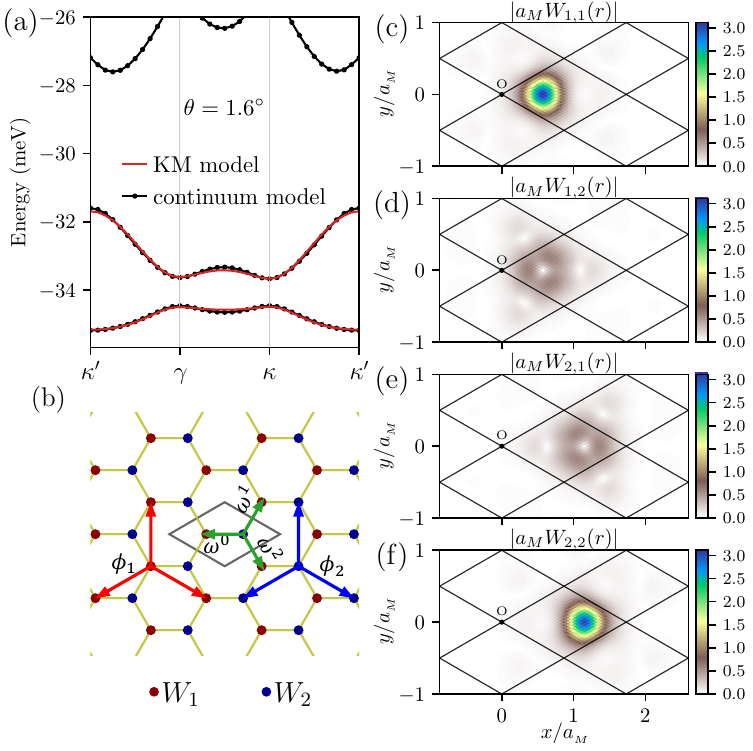}
	\centering
	\caption{\label{wannier}
		(a) Bosonic Kane-Mele tight-binding model dispersion (red solid line) 
		compared with the continuum model bands with $\theta=1.6^{\circ}$ and $eV_b-2\Delta E^0=4.5$ meV. 
		(b) Schematic illustration of the Kane-Mele tight-binding model (for valley $\tau=+K$)
		with nearest- (NN) and next-nearest-neighbor (NNN) hoppings. $\omega^n$ and $\phi_\alpha$ represent the 
		phase of the NN and NNN hopping processes, respectively, where $\omega=e^{i2\pi/3}$.
		The black diamond marks a real space moir\'e unit cell.
		(c)-(f) Amplitudes of the Wannier orbitals in real space. $W_{\alpha,l}$ is the $X_l$  component of the $W_{\alpha}$ orbital at the moir\'e site $\bm{R}=0$. The black grids mark the moir\'e unit cells. 
	}
\end{figure}

\emph{Bosonic Kane-Mele model}---\noindent
The topology and real space density \cite{SM} of 
the two lowest exciton moir\'e bands indicate that they can be described 
by an effective Kane-Mele lattice model with two orbitals (or sublattices)
in each unit cell.
{\colorrev To confirm this, we construct Wannier states for the two lowest bands
in regions where their total Chern number is zero;
outside these regions, additional bands and thereby Wannier orbitals must be included.}
The Wannier functions (for valley $+K$) at $\bm{R}=0$ are given by
\begin{align}
	|W_{\alpha}\rangle = \frac{1}{\sqrt{N}}\sum_{n=0,1}\sum_{\bm{Q}}F^{n}_{\alpha,\bm{Q}}|\Psi_{n,\bm{Q}}\rangle
\end{align}
for $\alpha=1,2$. $F_{\bm{Q}}$ is a unitary matrix for fixing the gauge of the wave function $|\Psi_{n,\bm{Q}}\rangle$.
We obtain $F_{\bm{Q}}$ by requiring that
$\sum_nF^{n}_{\alpha,\bm{Q}}|\Psi_{n,\bm{Q}}\rangle$ is maximally polarized to its $X_{l=\alpha}$ component
and is real at its center (where its amplitude peaks) in real space.
(See details in SM \cite{SM} and also Ref.~\cite{WuWannier}.)
Figures~\ref{wannier}(c)-\ref{wannier}(f) plot the $X_1$ and $X_2$ components of 
$W_{\alpha}(\bm{r})=[W_{\alpha, 1}(\bm{r}), W_{\alpha, 2}(\bm{r})]^T$ for site $\bm{R}=0$.
$W_{\alpha}(\bm{r})$ has dominant weight in its $W_{\alpha, \alpha}(\bm{r})$ component
and is centered around $\bm{r}=\bm{t}_{\alpha}$,
where $\bm{t}_{1}=(1/\sqrt{3},0)a_M$ and $\bm{t}_{2}=(2/\sqrt{3}, 0)a_M$ correspond to 
the $XM$ and $MX$ positions, respectively, in the moir\'e superlattice.
We construct the bosonic tight-binding model for the honeycomb lattice formed 
by the $W_{1}$ and $W_{2}$ orbitals as shown in Fig.~\ref{wannier}(b). 
The effective lattice Hamiltonian, keeping the nearest and next-nearest hopping terms, takes the form
\begin{align}
	\mathcal{H}^{+K}_{eff} =
	 &-\sum_{\alpha, i}\sum_{m=1}^3 t^{(2)}_{\alpha}e^{i\phi_{\alpha}} \hat{b}^{\dagger}_{\alpha, \bm{R}_i+\bm{a}_m}
	\hat{b}_{\alpha, \bm{R}_i} \notag\\
	&  -\sum_{\alpha, i}\sum_{m=1}^3 t^{(1)}e^{i\omega^{m-1}} \hat{b}^{\dagger}_{1, \bm{R}_i+\bm{\delta}_m}\hat{b}_{2, \bm{R}_i}+H.c.
	\label{KMham}
\end{align}
where $\hat{b}^{\dagger}_{\alpha, \bm{R}_i}$ is the bosonic creation operator for the $W_{\alpha,\bm{R}_i}$ state.
$\bm{a}_{1}=(\sqrt{3}/2,-1/2)a_M$ 
and $\bm{a}_m=(\hat{\mathcal{R}}_{2\pi/3})^{m-1}\bm{a}_1$ 
are real lattice vectors, and $\bm{\delta}_m=\{\bm{0}, -\bm{a}_3, \bm{a}_1\}$.
The first term is the intrasublattice hopping with amplitude $t^{(2)}_{\alpha}$ and phase $\phi_\alpha$.
$\phi_1=0$ and $\phi_2=\bm{\kappa}\cdot\bm{a}_1$ in the limit of zero interlayer hybridization, $T=0$, 
and gradually deviates from these values as $T$ is increased.
The second term is the nearest interorbital hopping term where $\omega=e^{i2\pi/3}$.
The bond-dependent phase seemingly breaks $C_3$ symmetry but, in fact, does not.
This is because $|W_1\rangle$ and $|W_2\rangle$ transform differently under $C_3$,
$C_3|W_{1,\bm{R}}\rangle=|W_{1,\bm{R}'}\rangle$, and $C_3|W_{2,\bm{R}}\rangle=\omega|W_{2,\bm{R}'}\rangle$
with $\bm{R}'+\bm{t}_\alpha=\hat{\mathcal{R}}_{2\pi/3}(\bm{R}+\bm{t}_\alpha)$ (See SM \cite{SM}).
$\mathcal{H}^{+K}_{\rm eff} $ is a generalized version of the Haldane Hamiltonian \cite{Haldane1988}
and, together with the $\mathcal{H}^{-K}_{\rm eff}$, forms the Kane-Mele (KM) model \cite{KaneMeleModel}.
The dispersion of the effective KM model agrees well with the continuum model band structure 
as illustrated in Fig.~\ref{wannier}(a), confirming the accuracy of truncating at the NNN hoppings.

\emph{Discussions and outlook}---\noindent
The  $tMX_2/M'X_2$ moir\'e heterostructure we propose realizes topological excitons with extended lifetimes, 
crucial for exploring strongly correlated many-body bosonic phases,
and establishes a new mechanism for inducing exciton topology that has not been explored before.
Most excitingly, it provides a promising route for realizing the first solid-state-based platform for the bosonic Kane-Mele model, 
which features topological flatbands and is, therefore, capable of simulating strongly correlated topological bosons.
It has been predicted that, in a Haldane-Bose-Hubbard model with topological flatbands, 
fractional quantum anomalous Hall states emerge at even denominator filling factors $\nu=1/2$ and $1/4$ \cite{Abelian}.
Interestingly, convincing numerical evidence of a stable bosonic non-Abelian quantum Hall state (without Landau level) is also found in such system at $\nu=1$ \cite{NonAbelian}.
Given the unprecedented tunability of TMD moir\'e superlattice, our proposal elevates the potential for uncovering these correlated topological phases, especially the non-Abelian quantum anomalous Hall state.

{ \colorrev
A hallmark of topological exciton bands is the presence of helical edge states inside the noninteracting topological gap, which can be probed by local scanning probes\cite{Feng2024} and optical reflection measurements \cite{HeinzScience} with moiré-scale spatial resolution. 
In the FQAHE phases predicted at finite exciton densities, 
bulk incompressibility manifests as jumps in the spectral peak energy 
upon increasing density. 
While the nature of their edge states remains largely unexplored, 
we anticipate several characteristics inferred from the FQHE case
\cite{FeldmanCh4,AshvinCorrelation,MontangeroCorrelation, JakschEdge, Kwan2022}, 
including algebraically decaying correlation functions, chiral edge channels, and fractionalization. The first two can be detected optically through spatial correlation and diffusion measurements. 
Electrical generation and control of excitons \cite{Xie2018,Ma2021,Nguyen2023,Joe2023}
also enable exciton electric and thermal transport measurements, 
potentially revealing the fractional nature of the edge states.
}


\begin{acknowledgments}
	{\em Acknowledgments}---\noindent	The authors acknowledge helpful discussions with Allan MacDonald, You Zhou, Jiabin Yu, Beini Gao, and Lifu Zhang. This work was supported by Laboratory for Physical Sciences.
	M. H. acknowledges support from the Army Research Office (ARO) under Grant No. W911NF-20-1-0232.
\end{acknowledgments}



\begin{thebibliography}{99}
\bibitem{footnote}
	Twisted WSe$_2$ and MoTe$_2$ homobilayers are both experimentally verified examples 
	which have a nontrivial topological interlayer hybridization. 
	We chose the tWSe$_2$ as an example because the common chalcogen specie of MoTe$_2$, 
	i.e., WTe$_2$,  does not have a stable $R$-type crystal structure.
\end{thebibliography}
\end{document}